\documentclass[12pt]{article}
\usepackage{amssymb,amsmath,epsfig}

\begin{document}

\title{\bf Behaviour of Charged Collapsing Fluids after Hydrostatic Equilibrium in $R^n$ Gravity}

\author{Hafiza Rizwana Kausar
\thanks{rizwa\_math@yahoo.com}\\\\
Centre for Applicable Mathematics \& Statistics,\\
UCP Business School, Faculty of Management Studies, \\ University of Central Punjab,
Lahore-Pakistan.}
\date{}
\maketitle

\begin{abstract}
The purpose of this paper is to study the transport equation and its
coupling with Maxwell equation in the framework of $R^n$ gravity.
Using Muller-Israel-Stewart theory for the conduction of dissipative
fluids, we analyze the temperature, heat flux, viscosity and thermal
conductivity in the scenario of relaxation time. All these
thermodynamical variables are appeared in the form of a single
factor whose influence is discussed on the evolution of relativistic
model for the heat conducting collapsing star.
 \end{abstract}

\section{Introduction}

The evolution of gravitational collapse and self-gravitating systems
has been widely discussed in General Relativity (GR). This type of study
usually based upon perturbing the system by changing its equilibrium
state. The tendency of the evolution of the object is studied as
soon as its departure from the equilibrium state. This method of
perturbation consists only linear terms by ignoring the quadratic
and higher order terms. In such cases, the evolution processes of
the self-gravitating systems may take place on the hydrostatic time
scale and quasi-static approximation could fail. Then, it is
necessary to study the evolution of the system immediately after its
departure from the equilibrium state on a time scale of the order of
relaxation times. The relaxation process may change the final
outcome of the gravitational collapse drastically. There are
particular cases of the collapsing spheres in the literature, where
relaxation time may cause the bounce or collapse of the evolving
system \cite{herrera}.

The applications of the electromagnetic field in astronomy and
astrophysics is an active research domain. A lot of work has been
devoted to discuss the collective effects of electromagnetic and
gravitational fields. For example,  the relativistic jets are a natural outcome of some of the most violent and spectacular astrophysical phenomena, such as the core collapse
of massive stars in gamma-ray bursts (GRBs) and the accretion onto
supermassive black holes in active galactic nuclei (AGN) {\cite{1508.02721}}. It is
generally accepted that these jets are powered electromagnetically
by the magnetized rotation of a central compact object, i.e., a
black hole or neutron star. The main source of power of AGN and GRB
jets is the rotational energy of the central black hole \cite{9,10}
and its accretion disk. The naturally occurring low mass density
and hence high magnetization of black-hole magnetospheres suggests
that the relativistic jets originate directly from the black-hole
ergosphere. As the plasma is attracted towards the compact object,
it is accelerated to the relativistic speeds and the in-falling
material typically forms an accretion disc around the compact object
\cite{Shakura and Sunyaev}. Plasma thermalization processes within
the accretion disc are thought to accelerate charged particles and
launch jets through shocks \cite{Das et al}.

The phenomenologies of gravitational collapse is of great interest
in modified gravity theories. In particular, $f(R)$ gravity is more
popular due to its straightforward generalization of GR and its
cosmological applications to accommodate the early \cite{31} or late
time \cite{32} acceleration of the universe. It is supposed that
$f(R)$ gravity can produce some kind of repulsive force similar to
that of dark energy in the Einstein gravity. Thus, the question
arises that whether such kind of repulsive effect in $f(R)$ gravity
can hinder the gravitational collapse and the formation of black
holes. Different aspects of gravitational collapse and black hole
formation for the spherically symmetric solution in $f(R)$ theory
have been explored \cite{30}, however, the dynamical and transport
process of forming black holes in $f(R)$ gravity through
gravitational collapse is not widely discussed. This work may
contribute to explore answers of such questions.

Many choices for the function $f(R)$ have appeared in the literature
which aimed to explain dark energy and accelerating universe \cite{revreport2}.
However, in this paper, we restrict ourself to a power law form
of $f(R)$ gravity, that is $f(R)=R^n$. This model has an important
physical meaning being determined by the presence of Noether
symmetries in the interaction Lagrangian \cite{20}. Earlier, this
model was constrained by  solar system test and to explain
accelerating universe \cite{revreport3}. However, to obtain analytical or qualitative
insight on exact solutions, it is perfectably acceptable model to
study $f(R)$ theory as a toy model. Also, the applications of this
model goes to the study of the dark matter by using spherically
symmetric solutions via Noether symmetries \cite{2 refree}. Such approach yield that $R^n$ type models are compatible with the spherically symmetry which has closed relevance with the Birkhoff theorem. The validity of this theorem is directly related to  the  physical properties of the self-gravitating system, e.g., stability and stationarity etc  \cite{1 refree}. 

In fact, the  relations between the fundamental plane parameters of
galaxies and the corrected Newtonian potential, coming from $R^n$,
can be found and justified from a physical point of view to fit the
observations \cite{1610.03336}.  Furthermore, the excellent
agreement among the theoretical and observed rotation curves and the
values of the stellar mass-to-light ratios with the predictions of
population synthesis models make us confident that $R^n$ gravity may
represent a good candidate to solve both the dark energy problem on
cosmological scales and the dark matter one on galactic scales with
the same value of the slope $n$ of the higher order gravity
Lagrangian \cite{0603522}.

In this paper, we discuss the transport equation of gravitational
collapse along with Maxwell source in $R^n$ gravity. We derive
general transport equation for $f(R)=R^n$ model and then fix $n=2$
as well to write the some results because the $R^2$ term could work
effectively at infrared scale. In the next section $\mathbf{2}$, the
modified Einstein field equations for the $R^n$ gravity combining
with Maxwell source are presented. In the section $\mathbf{3}$, we
will formulate the dynamical equations. The central problem, the
transport equation, is analysed in section $\mathbf{4}$ and  results
are provided in the last section.

\section{ Field Equations for $R^n$ Gravity and Maxwell Source}

The 4-dimensional $(\mu, \nu =0,1,2,3)$ action in $f(R)$ gravity
along with the Maxwell source and matter Lagrangian is defined as
\begin{equation}\label{action}
S=\frac{1}{2}\int
d^{4}x\sqrt{-g}\left(\frac{f(R)}{\kappa}-\frac{\digamma}{2\pi}\right)+\int
d^{4}xL_m(g_{\mu\nu},\Psi_m),
\end{equation}
where $\kappa$ stands for the coupling constant and
$\digamma=\frac{1}{4}F^{\mu\nu}F_{\mu\nu}$ is the Maxwell invariant
and $L_m$ is the Lagrangian for the matter source depending upon the
$g_{\mu\nu}$ and the matter field. The Maxwell equations in a
gravitational field enhance the gravity background  by the mass
energy relationship. Deriving the field equations by varying the
above action with respect to $g_{\mu\nu}$, we get the following set
of field equations
\begin{equation}\label{FE1}
f_RR_{\mu\nu}-\frac{1}{2}f(R)g_{\mu\nu}-\nabla_{\mu} \nabla_{\nu}f_R+ g_{\mu\nu}
\Box f_R=\kappa (T^m_{\mu\nu}+E_{\mu\nu}),\,
\end{equation}
where $T_{\mu\nu}^m= -\frac{2}{\sqrt{-g}}\frac{\delta L_m}{\delta
g^{\mu \nu}}$. Above equations can be written in the standard format
as follows
\begin{equation}\label{FE2}
G_{\mu\nu}=\kappa (T^D_{\mu\nu}+T^m_{\mu\nu}+E_{\mu\nu}),
\end{equation}
where the quantities on the right hand side are followed as
\begin{eqnarray}\label{Eff}
T_{\mu\nu}^{(D)}&=&\frac{1}{\kappa}\left(\frac{f-Rf_{,R}}{2}g_{\mu\nu}+\nabla_{\mu}
\nabla_{\nu}f_{,R} -g_{\mu\nu} \Box f_{,R}\right),\\\label{EM}
T^m_{\mu\nu}&=&(\rho+p)u_{\mu}u_{\beta}-pg_{\mu\nu}+
q_\mu u_\nu+q_{\nu}u_\mu,\\
E_{\mu\nu}&=&\frac{1}{4\pi}(-F^{\gamma}_{\mu}F_{\nu\gamma}
+\frac{1}{4}F^{\gamma\delta}F_{\gamma\delta}g_{\mu\nu}).
\end{eqnarray}
Here $q^u$ denote the heat flow vector satisfying $q^\mu u_\mu$, the
relation $F_{\mu\nu}=\Phi_{\nu,\mu}-\Phi_{\mu,\nu}$ is called the
field strength tensor and $\Phi_\mu$ is the electro magnetic tensor.
In terms of this strength tensor, the field equations for the
Maxwell source can be written as
\begin{eqnarray}
F^{\mu\nu}_{;\nu}&=&\mu_0J^\mu,\\
F_{[\mu\nu;\gamma]}&=&0,\\
J^\mu&=&\rho(t,r)V^\mu.
\end{eqnarray}
The quantities $J^\mu$, $\mu_0$, $V^\mu$  and $\rho$ are the four current, magnetic
permeability, the four velocity and the charge density respectively. In this paper,
we suppose that charge is at rest and hence the  magnetic field is zero, so that
\begin{equation}
\Phi_\mu=\Phi(t,r)\delta^{0}_{\mu}.
\end{equation}

We consider a spherically symmetric spacetime with general metric
components $A$, $B$, and $C$ as a function of time and radial
coordinates. This interior metric represent the matter source which
is undergoing dissipative process causing the gravitational
collapse. This interior matter is bounded by a spherical surface
$\Sigma$ and is given by
\begin{equation}\label{interior}
ds^2_-=A^2dt^{2}-B^2dr^{2}-C^2(t,r)(d\theta^{2}+\sin^2\theta
d\phi^{2}).
\end{equation}
For the metric exterior to the boundary surface, we consider a
spacetime represented in the form of a total charge $Q$ and a total
mass $M$ of the collapsing matter inside the $\Sigma$. This is given
by
  \begin{equation}\label{exterior}
ds^2_+=\left(1-\frac{2M(\nu)}{r}+\frac{Q^2}{r}\right)d\nu^2+2dr d\nu-r^2(d\theta^2+\sin^2\theta
d\phi^2).
\end{equation}
For the general interior spacetime, the Maxwell field
equations  will take the form
\begin{eqnarray}\label{EM1}
\frac{\partial^2\Phi}{\partial
r^2}-\left(\frac{B'}{B}+\frac{A'}{A}-\frac{2C'}{C}\right)\frac{\partial\Phi}{\partial
r}&=&4\pi\rho B^2A,\\\label{EM2}
\frac{\partial}{\partial t}\left(\frac{\partial\Phi}{\partial r}
\right)-\left(\frac{\dot{B}}{B}+\frac{\dot{A}}{A}-\frac{2\dot{C}}{C}\right)\frac{\partial\Phi}{\partial
r}&=&0.
\end{eqnarray}
The derivatives with respect to time and radius are denoted by dot
and prime respectively. Following the conservation law of four
current, i.e., $J^\mu;$ $_\mu=0$, we obtain the expression for the
charge and the electric field intensity per unit surface area as
following
\begin{eqnarray}
q(r)&=&2\pi\int^{r}_{0}\rho BC^2dr,\\
E(t,r)&=&\frac{q}{4\pi C^2}.
\end{eqnarray}

We consider background model $f(R)=R^n$, to formulate the field
equations using interior spacetime metric. The non-zero components
of the field equations are as follows
\begin{eqnarray}\nonumber
G_{00}&=&\frac{\kappa}{nR^{n-1}}\left[{\rho}A^{2}+\frac{A^2}{\kappa}\left\{\frac{(1-n)R^n}{2}
+\frac{n(n-1)[(n-2)R^{n-3}R'^2+R'']}{B^2}\right.\right.\\\nonumber
&+&\left(\frac{2\dot{C}}{C}-\frac{\dot{B}}{B}\right)
\frac{n(n-1)R^{n-2}\dot{R}}{A^2}+\left(\frac{2C'}{C}-\frac{B'}{B}\right)
\left.\left.\frac{n(n-1)R^{n-3}R'}{B^2}\right\}\right.\\\label{F00}&+&\left.2\pi
E^2\right],\\\nonumber
G_{01}&=&\frac{\kappa}{nR^{n-1}}\left[-qAB+\frac{1}{\kappa}\left(n(n-1)[(n-2)R^{n-3}R'\dot{R}+R^{n-2}\dot{R}']\right.\right.\\\label{F01}
&-&\left.\left.\frac{A'}{A}[n(n-1)R^{n-2}\dot{R}]-\frac{\dot{B}}{B}[n(n-1)R^{n-2}R']\right)\right],\\\nonumber
G_{11}&=&\frac{\kappa}{nR{n-1}}\left[p_rB^{2}-2\pi
E^2-\frac{B^2}{\kappa}\left\{\frac{(1-n)R^n}{2}+\frac{n(n-1)R^{n-2}\dot{R}}{A^2}\right.\right.\\\nonumber
&\times&\left.\left.\left(\frac{\dot{A}}{A}+\frac{2\dot{C}}{C}\right)-\left\{n(n-1)[(n-2)R{n-3}\dot{R}+R^{n-2}\ddot{R}]\right\}A^2\right.\right.\\\label{F11}
&+&\left.\left.\left(\frac{A'}{A}+\frac{2C'}{C}\right)\frac{n(n-1)R^{n-2}R'}{B^2}\right\}-2\pi
E^2\right],\\\nonumber
G_{22}&=&\frac{\kappa}{nR^{n-1}}\left[p_{\perp}C^2-\frac{C^2}{\kappa}
\left\{\frac{(1-n)R^n}{2}-\frac{n(n-1)[(n-2)R^{n-3}\dot{R}+R^{n-2}\ddot{R}]}{A^2}\right.\right.\\\nonumber
&+&\frac{n(n-1)[(n-2)R^{n-3}R'^2+R^{n-2}R''}{B^2}+\frac{nR^{n-1}}{A^2}\left(\frac{\dot{A}}{A}-\frac{\dot{B}}{B}+\frac{\dot{C}}{C}\right)\\\label{F22}
&+&\left.\left.\left(\frac{A'}{A}-\frac{B'}{B}+\frac{C'}{C}\right)\frac{n(n-1)R^{n-2}R'}{B^2}\right\}+2\pi
E^2\right].
\end{eqnarray}
To discuss the collapsing matter inside the star, the proper time,
proper radial derivatives and the collapsing velocity of the
dissipative fluid can be written as below
\begin{equation}\label{2.4}
D_{T}=\frac{1}{A}\frac{\partial}{\partial t},\quad
D_{C}=\frac{1}{C'}\frac{\partial}{\partial r} \quad U=D_{T}C=\frac{\dot{C}}{A},
\end{equation}
where the velocity always be considered negative to represent
collapse. The time derivative of the collapsing velocity, the
acceleration, $D_{T}U$, can be calculated using Eqs.(\ref{F11}) and
(\ref{2.4}) as follows
\begin{eqnarray}\nonumber
D_{T}U&=&\frac{A'}{AB}\tilde{E} -\frac{\kappa}{2nR^{n-1}}\left
[p-\frac{1}
{\kappa}\left\{-\frac{n(n-1)[(n-2)R^{n-3}\dot{R}+R^{n-2}\ddot{R}]}{A^2}\right.\right.\\\nonumber
&+&\left.\left.\frac{n(n-1)R^{n-2}\dot{R}}{A^2}\left(\frac{\dot{A}}{A}
+\frac{2\dot{C}}{C}\right)
+\frac{n(n-1)R^{n-2}R'}{B^2}\left(\frac{A'}{A}+\frac{2C'}{C}\right)\right.\right.\\\label{2.8}&+&\left.\left.\frac{(1-n)R^n}{2}\right\}\right].
\end{eqnarray}
The notation $\tilde{E}$ has been defined in terms of the Misner and
Sharp mass function $m$ as
\begin{equation}\label{24}
\tilde{E}=\left[1+U^{2}+\frac{2m}{C}\right]^{1/2}.
\end{equation}

\section{Transport Equation}

To study transport equation, first we need to formulate the dynamical equations of the collapsing fluid by using contracted Bianchi identities achieved by taking covariant derivative with respect to the four velocity $V_{\alpha}$ and four vector $\chi_\alpha=B^{-1}\delta^\alpha_1$ along the radial direction as $[\overset{(m)}{T^{\alpha\beta}}+\overset{{(D)}}
{T^{\alpha\beta}}+E^{\alpha\beta}]_{;\beta}V_{\alpha}=0$ and  $[\overset{(m)}{T^{\alpha\beta}}+\overset{{(D)}}{T^{\alpha\beta}}+T^{\alpha\beta}]_{;\beta}
\chi_{\alpha}=0$ respectively. The resulting lengthy equations obtained from these identities are given in appendix.
 Extracting the term $\frac{A'}{AB}(\rho+p)$ from Eq.(\ref{2.8}), we get
\begin{eqnarray}\nonumber
\frac{A'}{AB}(\rho+p)&=&\frac{(\rho+p)}{\tilde{E}}D_{T}U
+\frac{(\rho+p)\kappa}{2nR^{n-1}\tilde{E}}\left
[p-\frac{1}{\kappa}\left\{\frac{n(n-1)R^{n-2}\dot{R}}{A^2}
\right.\right.\\\nonumber &\times&\left.\left.\left
(\frac{\dot{A}}{A}+\frac{2\dot{C}}{C}\right)+\frac{(1-n)R^n}{2}
+\frac{n(n-1)R^{n-2}R'}{B^2}\left(\frac{A'}{A}+\frac{2C'}{C}\right)\right.\right.\\\label{2.9}
&-&\left.\left.\frac{n(n-1)[(n-2)R^{n-3}\dot{R}+R^{n-2}\ddot{R}]}{A^2}\right\}\right].
\end{eqnarray}
Replacing this term in Eq.(\ref{B2}), we get the following
expression for the acceleration of the collapsing fluid
\begin{eqnarray}\nonumber
&&(\rho+p)D_{T}U\\\nonumber&=&-(\rho+p)\left[\frac{\kappa
p_r}{2nR^{n-1}}-\frac{1}{2nR^{n-1}}\left\{\frac{(1-n)R^n}{2}-D_T\left(\frac{n(n-1)R^{n-2}
\dot{R}}{A}\right)\right.\right.\\\nonumber&+&\left.\left.\frac{nR^{n-1}D_TC}{C}
+\frac{\tilde{E}}{B}\left(\frac{A'}{A}+\frac{2C'}{C}\right)nR^{n-1}D_CR\right\}\right]-\tilde{E}\left[\left(\frac{D_TB}{B}-\frac{D_TC}{C}\right)\right.\\\nonumber
&\times&\left.2q-D_Tq\right]-\tilde{E}^2D_Cp
+\frac{\tilde{E}}{\kappa
B}S_{R^n},
\end{eqnarray}
where, we denote $S_{R^n}$ term as appeared purely due to $R^n$
gravity. Explicitly, it is written in the appendix.

To derive the  equation for the heat flux, we use the
M$\ddot{u}$ller-Israel-Stewart theory which help us to write thermal
conductivity in terms of a linear combination of various fluxes,
e.g., four velocity heat flux, etc. This theory has been conceived
in a series of papers by Israel and Stewart \cite{ Israel stewart}
followed by work of Muller \cite{Muller}. The study of the transport
equation obtained from this theory provide the information about the
the transfer of mass, heat and momentum during the matter collapse.
The equation is given by \begin{equation} \label{3.1} \tau
h^{\alpha\beta}u^{\gamma}q_{\beta;\gamma}+q^{\alpha}=-\eta
h^{\alpha\beta}(T_{,\beta}+a_{\beta}T)-\frac{1}{2}\eta
T^2\left(\frac{\tau u^{\beta}}{\eta T^2}\right)_{;\beta}q^{\alpha}.
\end{equation}
Here $h^{\alpha\beta}=g^{\alpha\beta}-u^{\alpha}u^{\beta}$ is the
projection tensor whereas the notation  $\eta$, $\tau$, $T$  and
$a_{\beta}T$ denotes the thermal conductivity, the relaxation time,
the temperature and the Tolman inertial term with
$a_{\alpha}=u_{\alpha;\beta}u^{\beta}$ being the acceleration
respectively. The non-zero and independent component of the above
equation is given by
\begin{equation}\label{3.2}
\tau \dot{q}=-qA-\frac{1}{2}\eta qT^2\left(\frac{\tau}{\eta
T^2}\right)^{\cdot}-\frac{1}{2}\tau
q\left(\frac{\dot{B}}{B}+2\frac{\dot{C}}{C}\right) +\frac{\eta A^2
}{B}\left(\frac{T}{A}\right)'.
\end{equation}
Eliminating the expression for the quantity, $\frac{A'}{A}$, from
Eq.(\ref{2.9}) and then substituting it in Eq.(\ref{24}), we obtain
\begin{eqnarray}\nonumber
D_{T}q&=&-\frac{\eta T^2q}{2\tau}D_{T}\left(\frac{\tau }{\eta
T^2}\right)-\frac{q}{2}\left(\frac{D_TB}{B}+\frac{2D_TC}{C}+\frac{1}{\tau}\right)+\frac{\eta
\tilde{E}}{\tau }\left(D_{C}T\right.\\\nonumber
&-&\left.\frac{D_{T}U}{\tau \tilde{E}^2}\right)-\frac{\eta T}{\tau
\tilde{E}}\left[\frac{\kappa
p_r}{2nR^{n-1}}-\frac{1}{2nR^{n-1}}\left\{-D_T\left(\frac{n(n-1)R^{n-2}
\dot{R}}{A}\right)\right.\right.\\\nonumber&+&\left.\left.\frac{nR^{n-1}D_TC}{C}
+\frac{(1-n)R^n}{2}+\frac{\tilde{E}}{B}\left(\frac{A'}{A}+\frac{2C'}{C}\right)nR^{n-1}D_CR\right\}\right].\\\label{3.3}
\end{eqnarray}
To see the effects of the heat flux on the dissipative process of the collapsing
fluid, we use this version of the transport equation into  Eq.(\ref{3.7})
and fix $n=2$ to obtain
\begin{eqnarray}\nonumber
&&(\rho+p)\left[1-\frac{\eta T}{\tau(\rho+p)}\right]D_{T}U=
-\left[1-\frac{\eta T}{\tau(\rho+p)}\right]\left(\frac{\rho+p}{4R}\right)\left[\frac{\kappa
p}{4R}\right.\\\nonumber&+&\left.\left.\frac{1}{4R}\left\{D_T\left(\frac{2\dot{R}}{A}\right)\right.+\frac{R^2}{2}-\frac{2RD_TC}{C}
-\frac{\tilde{2RE}}{B}\left(\frac{A'}{A}+\frac{2C'}{C}\right)D_CR\right\}\right]\\\nonumber
&+&\tilde{E}\left[\frac{\eta
T^2q}{2\tau}D_T\left(\frac{\tau}{\eta
T^2}\right)-\tilde{E}^2D_Cp\frac{q}{2}\left(\frac{D_TB}{B}
+\frac{2D_TC}{C}+\frac{1}{\tau}\right)
+\frac{\eta\tilde{E}}{\tau}D_CT\right]\\\label{TE}
&+&\frac{\tilde{E}}{\kappa
B}S_{R^2}.
\end{eqnarray}
This equation yield the energy transport in a star. There are three
ways of energy transfer from hot to cold layers of the star, i.e.,
conduction, radiation and convection. Usually, photons carry energy
from the hot interior core of a collapsing  star to the outer cold
space. If photons/radiation unable to transfer total energy of the
hot interior star to the the surface of the star, then the method of
convention is used to process energy transfer. In the method of
convention, hotter gases come to the upper levels of the star
surfaces to radiate their energy and meanwhile cooler gases sink
towards the hot interior to collect energy. The third way of the
transport energy is the conduction method in which each atom
transfer its energy to its neighbouring atoms, however, this method
is usually ignored due to its low efficiency.

\section{Discussion and Results}

In this paper, we have discussed the dynamics of the dissipative
fluid after its departure from the hydrostatic equilibrium by using
transport equation of a radiating charged fluid. We have adopted the power-law version of  $f(R)$ theory 
which  could fit well the observations  and  encourages further investigation on $R^n$ gravity from
both  theoretical and observational points of view \cite{0703243}. The field equation
have been derived for $R^n$ theory and Maxwell source.
The conservation equations for the  matter yielded two
types of dynamical equations. These dynamical equations give the
information about the dynamics of the collapsing fluid and also
further used in the transport equation for
M$\ddot{u}$ller-Israel-Stewart theory of dissipative fluids to get
hydrostatic equilibrium evolution equation. 

It is found that the resulting evolution equation (\ref{TE})
critically depend on a factor composed of thermodynamic variables.
On the left hand side of this equation, the term $D_TU$ is the
acceleration whereas the product term  $(\rho+p)$ is the inertial
mass density. Thus  by Newton's law, the right hand side term
represent the gravitational force term along with the repulsive term
$S_{R^2}$. To interpret, we suppose that gravitational force term
overcome the effect of repulsive term then we see that both sides of
the equation  are affected by the factor $1-\frac{\eta
T}{\tau(\rho+p)}$. Also, the same factor is appearing on the right
hand side of the equation and hence represent the consistency of the
equivalence principle. If we denote this factor by $\beta$, then we may have the following possibilities. 

(i)  If $0<\beta<1$, then inertia of heat causes a decrease in
inertial and gravitational mass densities due to fractional
factor. If the evolution proceeds in such a way that
$\beta\rightarrow1$, then the effective inertial mass density of
the fluid element approaches to zero. For a very small value of
the relaxation time at present time, we may speculate that
$\beta$ may increase substantially in a pre-supernovae event. In
fact, at the last stages of a massive star evolution, the decrease
of inertial densities would prevent the propagation of photons and
neutrino \cite{W}. 

(ii) If $\beta\rightarrow0$, then there is no effect on the
inertial mass density and gravitational force. In addition, this
case may lead to the fact that $\eta T\rightarrow0$. If this
happens, then the core becomes degenerate, starts to cool and the
star must becomes a white dwarf. This case may be fitted to the small bodies such as Saturn which is 
is stable against  the gravitational collapse. 
If we gave Saturn a slight squeeze, both the
 gravitational force and the pressure within its core would increase. 
 The gravitational force would rise simply as the inverse-square of the radius, 
 but the force of the pressure would rise faster than the inverse-square of the radius. 
This imbalance of forces would cause Saturn to expand back to its equilibrium radius  regardless of how cold Saturn grows. 

(iii) If $\beta>1$, then it changes the sign and hence the
gravitational force term becomes positive implying reversal of
collapse. Consequently, this case may stop the collapse and make
the star explodes. If this does not happen, the collapse would
lead to the region of instability. This mechanism is assumed to
cause type II supernovae.

(iv)  The case when $\beta=1$ we get the critical point during
gravitational collapse. In this case, the force terms on the right
hand side of the evolution equation will also be zero and we are
left with a constraint equation  as
\begin{eqnarray}\nonumber
&&\frac{\eta
T^2q}{2\tau}D_T\left(\frac{\tau}{\eta
T^2}\right)-\tilde{E}^2D_Cp\frac{q}{2}\left(\frac{D_TB}{B}
+\frac{2D_TC}{C}+\frac{1}{\tau}\right)
+\frac{\eta\tilde{E}}{\tau}D_CT\\\label{TE}
&&=\frac{\tilde{E}}{\kappa
B}S_{R^2}.
\end{eqnarray}
This equation represents the dissipative regime of the collapsing
sphere immediately after its departure from the equilibrium state on
a time scale of the relaxation time $\tau$. If we suppose that
before hydrostatic equilibrium, there is no dissipation then all
terms on the left hand side will vanish due to the vanishing of $q$
and $\eta$. At the moment of hydrostatic equilibrium, the relaxation
time influences the evolution process. Some particular values of
the relaxation time may cause the bounce or collapse of the sphere
\cite{Peirls, Harwit}.

If the inertial mass density $(\rho+p)$ will be zero for the perfect
fluid case, then the discussion will be same as for the factor
$\frac{\eta T}{\tau(\rho+p)}$ when it approaches to $1$. It is
mentioned that we are evaluating the system immediately after its
leaving from the equilibrium state, hence the physical meaning of
thermodynamical variables is hard to interpret numerically, however,  as a guess, the values may be as  are $[\eta]\approx 10^{37},~[T]\approx 10^{13},~[\tau]\approx 10^{-4},~ [\rho]\approx
10 ^{12}$ \cite{W} . In general, the obtained results represent general self-gravitating
dissipative fluid model. 

\section{Appendix}

For the model $f(R)=R^n$, the derivatives of $\frac{df}{dR}=f_R$ used in the field equation are given by
\begin{eqnarray}
\dot{f}_R&=&n(n-1)R^{n-2}\dot{R}\\
f'_R&=&n(n-1)R^{n-2}R'\\
\ddot{f}_R&=&n(n-1)[(n-2)R^{n-3}\dot{R}+R^{n-2}\ddot{R}]\\
\dot{f'}_R&=&n(n-1)[(n-2)R^{n-3}R'\dot{R}+R^(n-2)\dot{R}']\\
{f''}_R&=&n(n-1)[(n-2)R^{n-3}R'^2+R^{n-2}R'']
\end{eqnarray}

First dynamical equation, $[\overset{(m)}{T^{\alpha\beta}}+\overset{{(D)}}
{T^{\alpha\beta}}+E^{\alpha\beta}]_{;\beta}V_{\alpha}=0$ :

\begin{eqnarray}\nonumber
&&\frac{\dot{\rho}}{A}+\frac{q'}{B}+\frac{q}{B}\left(\frac{A'}{A}+\frac{2C'}{C}\right)+\frac{(\rho+p)}{A}\left(\frac{\dot{B}}{B}
+\frac{2\dot{C}}{C}\right)+\frac{A}{\kappa}\left[\frac{1}{A^2B^2}\right.\\\nonumber
&\times&\left.\left\{n(n-1)(n-2)R^{n-3}R'\dot{R}+R^{n-2}\dot{R}'-\frac{A'}{A}{n(n-1)R^{n-2}}\dot{R}\right.\right.\\\nonumber
&-&\left.\left.\frac{\dot{B}}{B}{n(n-1)R^{n-2}R'}\right\}\right]_{,1}+\frac{A}{\kappa}\left\{\frac{n(n-1)}{A^2B^2}
(n-2)R^{n-3}R'^2\right.\\\nonumber
&-&\left.\frac{n(n-1)R^{n-2}\dot{R}}{A^2}\left(\frac{\dot{B}}{B}-\frac{2\dot{C}}{C}\right)
-\frac{n(n-1)R^{n-2}R'}{B^2}\left(\frac{B'}{B}
-\frac{2C'}{C}\right)\right.\\\nonumber
&+&\left.\frac{(1-n)R^n}{2A^2}\right\}_{,0}+\frac{\dot{A}}{\kappa
A^2}\left\{\frac{(1-n)R^n}{2A^2}+\frac{n(n-1)}{A^2B^2}(n-2)R^{n-3}R'^2\right.\\\nonumber
&-&\left.\frac{n(n-1)R^{n-2}\dot{R}}{A^2}\left(\frac{\dot{B}}{B}-\frac{2\dot{C}}{C}\right)
-\frac{n(n-1)R^{n-2}R'}{B^2}\left(\frac{B'}{B}-\frac{2C'}{C}\right)\right\}
\end{eqnarray}
 \begin{eqnarray}\nonumber
&+&\frac{\dot{B}}{\kappa
AB}\left\{\frac{n(n-1)}{B^2}[(n-2)R^{n-3}R'^2+R^{n-2}R'']-\frac{n(n-1)R^{n-2}\dot{R}}{A^2}\right.\\\nonumber
&\times&\left.\left(\frac{\dot{A}}{A}+\frac{\dot{B}}{B}\right)
+\frac{n(n-1)}{A^2}[(n-2)R^{n-3}\dot{R}+R^{n-2}\ddot{R}]
-\left(\frac{A'}{A} +\frac{B'}{B}\right)\right.\\\nonumber
&\times&\left.\frac{n(n-1)R^{n-2}R'}{B^2}\right\}
+\frac{2\dot{C}}{\kappa
AC}\left\{\frac{n(n-1)}{A^2}[(n-2)R^{n-3}\dot{R}+R^{n-2}\ddot{R}]
\right.\\\nonumber
 &+&\left. \frac{n(n-1)R^{n-2}\dot{R}}{A^2}
\left(\frac{\dot{C}}{C}-\frac{\dot{A}}{A}\right)-\frac{n(n-1)R^{n-2}R'}{B^2}\left(\frac{A'}{A}
-\frac{C'}{C}\right)\right\}\\\nonumber
 &+&\frac{1}{\kappa
AB^2}\left(n(n-1)[(n-2)R^{n-3}R'\dot{R}+R^{n-2}\dot{R}']
-n(n-1)R^{n-2}\dot{R}\frac{A'}{A}\right.\\\label{19} &-&\left.
{n(n-1)R^{n-2}R'}\frac{\dot{B}}{B}\right)\left(\frac{2A'}{A}+\frac{B'}{B}+\frac{C'}{C}\right)=0.
\end{eqnarray}

Second dynamical equation,
$[\overset{(m)}{T^{\alpha\beta}}+\overset{{(D)}}{T^{\alpha\beta}}+T^{\alpha\beta}]_{;\beta}
\chi_{\alpha}=0$:

\begin{eqnarray}\nonumber
&&\frac{p'}{B}+\frac{\dot{q}}{A}+\frac{2q}{A}\left(\frac{\dot{B}}{B}
+\frac{\dot{C}}{C}\right)+\frac{A'}{AB}(\rho+p)-\frac{B}{\kappa}\left[\left\{\frac{n(n-1)}{A^2B^2}\right.\right.\\\nonumber
&\times&\left.[(n-2)R^{n-3}R'\dot{R}+R^{n-2}\dot{R'}]-n(n-1)R^{n-2}\left(\frac{A'}{A}\dot{R}+\frac{\dot{B}}{B}R'\right)\right\}_{,0}\\\nonumber
&+&\left\{\frac{(1-n)R^n}{2B^2}-\frac{n(n-1)}{A^2}[(n-2)R^{n-3}\dot{R}+R^{n-2}\ddot{R}]+\left(\frac{\dot{A}}{A}+\frac{2\dot{C}}{C}\right)\right.\\\nonumber
&\times&\left.\frac{n(n-1)R^{n-2}\dot{R}}{A^2}+\frac{B^2}{n(n-1)R^{n-2}R'}\left(\frac{A'}{A}
+\frac{2C'}{C}\right)\right\}_{,1}+\frac{A'}{AB^2}\\\nonumber
&\times&\left\{\frac{n(n-1)}{A^2}[(n-2)R^{n-3}\dot{R}+R^{n-2}\ddot{R}]-\frac{n(n-1)R^{n-2}\dot{R}}{A^2}+
\left(\frac{\dot{A}}{A}+\frac{\dot{B}}{B}\right)\right.\\\nonumber
&\times&\left.\frac{n(n-1)}{B^2}[(n-2)R^{n-3}R'^2+R^{n-2}R'']\right.
-\left.\frac{n(n-1)R^{n-2}R'}{B^2}\left(\frac{A'}{A}-\frac{B'}{B}\right)\right\}
\end{eqnarray}
\begin{eqnarray}\nonumber
&+&\frac{2B'}{B^3}\left(\frac{A'}{A}-\frac{C'}{C}\right)\left\{\frac{(1-n)R^n}{2A^2}+\frac{n(n-1)[(n-2)R^{n-3}\dot{R}+R^{n-2}\ddot{R}]}{A^2}\right.\\\nonumber
&+&\frac{A'}{A}{n(n-1)R^{n-2}\dot{R}}-\frac{\dot{B}}{B}{n(n-1)R^{n-2}R'}-\frac
{n(n-1)R^{n-2}\dot{R}}{A^2}\left(\frac{\dot{A}}{A}+\frac{\dot{B}}{B}\right)\\\nonumber
&-&\left.\frac{n(n-1)R^{n-2}R'}{B^2}\left(\frac{A'}{A}+\frac{B'}{B}\right)\right\}+\frac{2\dot{C}}{CA^2}\left\{\frac{n(n-1)}{A^2}[(n-2)R^{n-3}\dot{R}
\right.
\\\nonumber&+&\left.R^{n-2}\ddot{R}]+\frac{n(n-1)R^{n-2}\dot{R}}{A^2}\left(\frac{\dot{C}}{C}-\frac{\dot{A}}{A}\right.\right)
-\left.\frac{n(n-1)R^{n-2}R'}{B^2}\right\}+\frac{1}{A^2B^2}\\\nonumber&\times&\left.[n(n-1)(n-2)R^{n-3}R'\dot{R}+R^{n-2}\dot{R}']-\left(\frac{\dot{A}}{A}+\frac{3\dot{B}}{B}
+\frac{2\dot{C}}{C}\right)
\right]=0.\\\label{B2}
\end{eqnarray}
\begin{eqnarray}\nonumber
S_{R^n}&=&\left[\left(\frac{D_TA}{A}-\frac{D_TB}{B}-\frac{2D_TC}{C}\right)\left\{n(n-1)\right.\right.\\\nonumber&\times&\left.
D_TR^{n-2}R'-\frac{A'}{A}nR^{n-1}D_T\left(\frac{n(n-1)R^{n-2}R'D_TB}{B}\right)\right\}+\left\{2nR^{n-1}\right.\\\nonumber
&\times&\left.\frac{D_TC}{C}-D_T\left(\frac{n(n-1)R^{n-2}\dot{R}}{A}\right)+\frac{\tilde{E}}{B}\left(\frac{A'}{A}
+\frac{2C'}{C}\right)nR^{n-1}D_CR\right\}_{,1}\\\nonumber
&+&\frac{A'}{A}\left\{D_T\left(\frac{n(n-1)R^{n-2}\dot{R}}{A}\right)+\frac{n(n-1)}{B^2}[(n-2)R^{n-3}R'^2+R^{n-2}R'']\right.\\\nonumber
&-&\left.\frac{n(n-1)R^{n-2}R'}{B^2}\left(\frac{A'}{A}+\frac{B'}{B}\right)-\frac{nR^{n-1}D_TBD_T}{B}\right\}+\frac{2C'}{C}
\left\{\frac{n(n-1)}{B^2}\right.\\\nonumber
&\times&\left.[(n-2)R^{n-3}R'^2+R^{n-2}R'']-nR^{n-1}D_T\left(\frac{D_TB}{B}+\frac{D_TC}{C}\right)\right.-\frac{n(n-1)}{B^2}\\\nonumber
&-&\left.R^{n-2}R'\left(\frac{B'}{B}+\frac{C'}{C}\right)\right\}+\frac{1}{A^2}\left\{n(n-1)[(n-2)R^{n-3}R'\dot{R}+R^{n-2}\dot{R}']\right.\\\label{3.7}
&-&\left.\left.n(n-1)R^{n-2}\left(\dot{R}\frac{A'}{A}-R'\frac{\dot{B}}{B}\right)\right\}_{,0}\right].
\end{eqnarray}

\begin{eqnarray}\nonumber
S_{R^2}&=&\left[\left(\frac{D_TA}{A}-\frac{D_TB}{B}-\frac{2D_TC}{C}\right)\left\{2D_TR'-\frac{A'}{A}2RD_T\left(\frac{2R'D_TB}{B}\right)\right\}\right.\\\nonumber&+&
\left\{4R\frac{D_TC}{C}+2RD_CR\frac{\tilde{E}}{B}\left(\frac{A'}{A}
+\frac{2C'}{C}\right)-D_T\left(\frac{2\dot{R}}{A}\right)\right\}_{,1}\\\nonumber
&+&
\frac{A'}{A}\left\{D_T\left(\frac{2\dot{R}}{A}\right)+\frac{2R''}{B^2}-\frac{2R'}{B^2}\left(\frac{A'}{A}+\frac{B'}{B}\right)-\frac{2RD_TBD_T}{B}\right\}\\\nonumber&-&\frac{2C'}{C}
\left\{\frac{2R''}{B^2}-2D_T\left(\frac{D_TB}{B}+\frac{D_TC}{C}\right)-R'\left(\frac{B'}{B}+\frac{C'}{C}\right)\right\}\\\label{3.9}
&-&\left.\frac{2}{A^2}\left\{\left(\dot{R}\frac{A'}{A}-R'\frac{\dot{B}}{B}\right)+\frac{\dot{R}'}{A^2}\right\}_{,0}\right].
\end{eqnarray}

\end{document}